\documentclass[conference]{IEEEtran}
\IEEEoverridecommandlockouts
\usepackage{cite}
\usepackage{amsmath,amssymb,amsfonts}
\usepackage{algorithmic}
\usepackage{graphicx}
\usepackage{textcomp}
\usepackage{xcolor}
\usepackage{float}
\usepackage{hyperref}

\usepackage{float,subfig}
\usepackage{tikz}
\usetikzlibrary{shapes,arrows,calc}
\usetikzlibrary{arrows.meta}
\usetikzlibrary{decorations.pathreplacing}
\usetikzlibrary{shapes.geometric}
\usetikzlibrary{shadows}
\usepackage{listings}
\usepackage{graphicx}
\usepackage{adjustbox}

\usepackage{verbatim}

\usepackage{array,colortbl,multirow,multicol,booktabs,ctable}

\definecolor{delim}{RGB}{0,0,0}
\definecolor{numb}{RGB}{0, 0, 0}
\definecolor{string}{rgb}{0.70,0.00,0.50}
\lstdefinelanguage{json}{
    numbers=left,
    numberstyle=\small,
    frame=single,
    rulecolor=\color{black},
    showspaces=false,
    showtabs=false,
    breaklines=true,
    postbreak=\raisebox{0ex}[0ex][0ex]{\ensuremath{\color{gray}\hookrightarrow\space}},
    breakatwhitespace=true,
    basicstyle=\ttfamily\small,
    upquote=true,
    morestring=[b]",
    stringstyle=\color{string},
    literate=
     *{0}{{{\color{numb}0}}}{1}
      {1}{{{\color{numb}1}}}{1}
      {2}{{{\color{numb}2}}}{1}
      {3}{{{\color{numb}3}}}{1}
      {4}{{{\color{numb}4}}}{1}
      {5}{{{\color{numb}5}}}{1}
      {6}{{{\color{numb}6}}}{1}
      {7}{{{\color{numb}7}}}{1}
      {8}{{{\color{numb}8}}}{1}
      {9}{{{\color{numb}9}}}{1}
      {\{}{{{\color{delim}{\{}}}}{1}
      {\}}{{{\color{delim}{\}}}}}{1}
      {[}{{{\color{delim}{[}}}}{1}
      {]}{{{\color{delim}{]}}}}{1},
}

\definecolor{blueLine}{RGB}{57,106,177}
\definecolor{blueFill}{RGB}{114,147,203}
\definecolor{redLine}{RGB}{204,37,41}
\definecolor{greenline}{RGB}{0,250,0}
\definecolor{blackLine}{RGB}{0,0,0}
\definecolor{goldLine}{RGB}{160,82,45}
\definecolor{goodGreen}{RGB}{213,232,212}
\definecolor{goodGreenBorder}{RGB}{150,192,129}
\definecolor{goodBlue}{RGB}{218,232,252}
\definecolor{goodBlueBorder}{RGB}{144,170,207}
\definecolor{goodPink}{RGB}{248,206,204}
\definecolor{goodPinkBorder}{RGB}{200,114,111}
\definecolor{airforceblue}{rgb}{0.36, 0.54, 0.66}
\definecolor{aquamarine}{rgb}{135, 206, 255}
\definecolor{deepskyblue}{rgb}{0.0, 0.75, 1.0}
\definecolor{persianblue}{rgb}{0.11, 0.22, 0.73}
\definecolor{aliceblue}{rgb}{0.94, 0.97, 1.0}

\def\BibTeX{{\rm B\kern-.05em{\sc i\kern-.025em b}\kern-.08em
    T\kern-.1667em\lower.7ex\hbox{E}\kern-.125emX}}

\IEEEoverridecommandlockouts
\begin{document}

\title{Anvil: An integration of artificial intelligence, sampling techniques, and a combined CAD-CFD tool\\

}

\author{
\IEEEauthorblockN{Harsh Vardhan\IEEEauthorrefmark{1}, Umesh Timalsina\IEEEauthorrefmark{1}, Michael Sandborn\IEEEauthorrefmark{1}, David Hyde\IEEEauthorrefmark{1}, Peter Volgyesi\IEEEauthorrefmark{1}, Janos Sztipanovits\IEEEauthorrefmark{1}}
\IEEEauthorblockA{\IEEEauthorrefmark{1} Vanderbilt University, Nashville, TN, USA \\
Email: \{harsh.vardhan, umesh.timalsina, michael.sandborn, david.hyde.1, peter.volgyesi, janos.sztipanovits\}@vanderbilt.edu}
\thanks{Manuscript accepted at 2024 6th Workshop on Design Automation for CPS and IoT (DESTION 2024), May 2024, Hongkong  and in proceeding to published in IEEE Xplore}
}

\maketitle
\begin{abstract}
In this work, we introduce an open-source integrated CAD-CFD tool, \textit{Anvil}, which combines FreeCAD for CAD modeling and OpenFOAM for CFD analysis, along with an AI-based optimization method (Bayesian optimization) and other sampling algorithms. \textit{Anvil} serves as a scientific machine learning tool for shape optimization in three modes: data generation, CFD evaluation, and shape optimization. In data generation mode, it automatically runs CFD evaluations and generates data for training a surrogate model. In optimization mode, it searches for the optimal design under given requirements and optimization metrics. In CFD mode, a single CAD file can be evaluated with a single OpenFOAM run. To use \textit{Anvil}, experimenters provide a JSON configuration file and a parametric CAD seed design. \textit{Anvil} can be used to study solid-fluid dynamics for any subsonic flow conditions and has been demonstrated in various simulation and optimization use cases. The open-source code for the tool, installation process, artifacts (such as CAD seed designs and example STL models), experimentation results, and detailed documentation can be found at \url{https://github.com/symbench/Anvil}.
\end{abstract}

\begin{IEEEkeywords}
Computer-Aided Design (CAD), Computational Fluid Dynamics (CFD), Bayesian Optimization (BO), Surrogate Modeling, Design Optimization.
\end{IEEEkeywords}

\section{Introduction}
Design optimization, which encompasses shape optimization, is a widely explored and essential task in computational physics~\cite{WANG2007395,HAZRA200546,VIQUERAT2021110080,ALEXANDROV2005121,AMSTUTZ2006573,ALLAIRE2004363}. Shape optimization involves modifying the geometry of a given topological architecture to achieve specific objectives by refining and improving the shape of an object. Modifications made to a shape under study might entail minimizing stress concentrations, maximizing strength tolerance on specified axes, or optimizing fluid flow characteristics. 
Our focus is on shape optimization design problems in engineering that require CAD modeling and fluid simulation to analyze a design.  

On the optimization front, recent advancements in artificial intelligence (AI) have opened up exciting possibilities for these algorithms' use as sample-efficient optimizers \cite{vardhan2023sample}. The fusion of traditional CAD-CFD toolchains with AI has the potential to revolutionize the way engineering simulations and design processes are conducted.

In the context of this type of optimization problem, we note two main challenges: 
\begin{enumerate}
\item \textbf{Lack of open-source, easy-to-use automated CAD-CFD evaluation tools}: The shape optimization process is hindered by the manual and disjointed use of CAD for design modifications and CFD for fluid dynamics analysis, necessitating manual file handling (e.g., creating and passing STL files between two programs). Seamless integration is essential for automating the design evaluation process, reducing manual intervention.
Despite the existence of integrated CAD-CFD commercial solutions, their high license fees often make them an infeasible option for researchers. 
     
\item \textbf{Time-consuming and computationally expensive evaluation process}: The evaluation process through CAD-CFD pipelines is notably time-consuming and computationally intensive, with simulations extending from minutes to days. This inefficiency is exacerbated in exhaustive search methods, particularly in complex design spaces. AI-based optimization methods offer a solution with their sample efficiency and superior convergence properties, underscoring the need for integrating these methods with CAD-CFD tools to enhance practicality and performance.   
\end{enumerate}
To address these challenges, we present an open-source, highly automated, and algorithmically efficient software tool, \textit{Anvil}, for CAD-CFD design optimization problems.  Our tool integrates existing open-source CAD and CFD tools and wraps this integrated functionality within a Python-based sampling and optimization framework.

The remainder of the manuscript is structured as follows.
After a discussion of related work in Section \ref{sec:related_work},
Sections \ref{sec:toolchain} presents the software architecture of \textit{Anvil}, including the standard for creating parametric CAD designs, numerical settings, mathematical models used for CFD simulation, and implementation details of the tool.
In Section \ref{sec:casestudy}, we show experiments conducted on various designs. No one tool is complete; accordingly, in Section \ref{sec:limitation}, we elucidate some of \textit{Anvil}'s limitations.
Concluding remarks and possible future directions are provided in Section \ref{sec:conclusion}.

\section{Related Work}
\label{sec:related_work}
The development of integrated CAD-CFD tools for shape optimization has been a focus in commercial tools like PTC Creo and AutoDesk Fusion360. However, the open-source equivalent of this capability is not readily available.   Previous works have utilized Bayesian optimization in loop with computational fluid dynamics (CFD) for design optimization \cite{vardhan2023sample,garcia2008cfd,griaznov1999cfd,vardhan2023constrained}, but these works did not make their entire toolchains available. Moreover, these solutions generally require manual intervention during  optimization process , which is a bottle neck. The design evaluation tool provided in \cite{vardhan2023sample} is only demonstrated for Myring hull design optimization in the context of unmanned underwater vehicles (UUVs). 
On the learning front, most traditional works explored the use of genetic algorithms combined with CFD simulations to optimize shapes \cite{marco1999multi}. Although effective in identifying optimal designs, genetic algorithms can be computationally expensive and may not always converge to the global optimum~\cite{vardhan2023sample}. In recent times, other learning models like random forest~\cite{vardhan2021machine}, deep learning \cite{vardhan2022deep,vardhan2022data}, and Bayesian optimization \cite{frazier2018tutorial,lepird2015bayesian} are gaining more attention due to their performance drastically exceeding classical approaches. 
The application of Bayesian optimization for shape optimization in fluid dynamics has been less common, with recent studies such as  \cite{morita2022applying,vardhan2023sample} demonstrating its potential in optimizing complex geometries with a limited number of simulations \cite{frazier2016bayesian}. However, these approaches have not been fully integrated into a user-friendly tool that combines CAD modeling and CFD analysis.

\section{Integration Architecture}
\label{sec:toolchain}
In this section, we provide a high-level overview of \textit{Anvil}. In computer-aided engineering (CAE), for shape optimization problems, computer-aided design (CAD) and computational fluid dynamics (CFD) are two essential technologies. CAD allows designers to create and modify models of solid objects, while CFD enables the analysis and simulation of fluid flow and related phenomena. The primary goal of CAD-CFD software architecture is to bridge the gap between the geometric design of objects and understanding objects' interactions with fluids. 

At its core, \textit{Anvil} comprises a unified software platform that integrates CAD and CFD tools, facilitating automated and accurate analysis of fluid behavior within complex geometries while also allowing flexibility to create, modify, and analyze 3D models of their designs. 
We then build a Python toolchain around these integrated software components, where the key advantage is the ability to automatically generate design variations, conduct CFD evaluation, and perform parametric optimizations. Accordingly, our architecture empowers users to incorporate optimization methods or automatic sampling methods to optimize their designs or generate data.

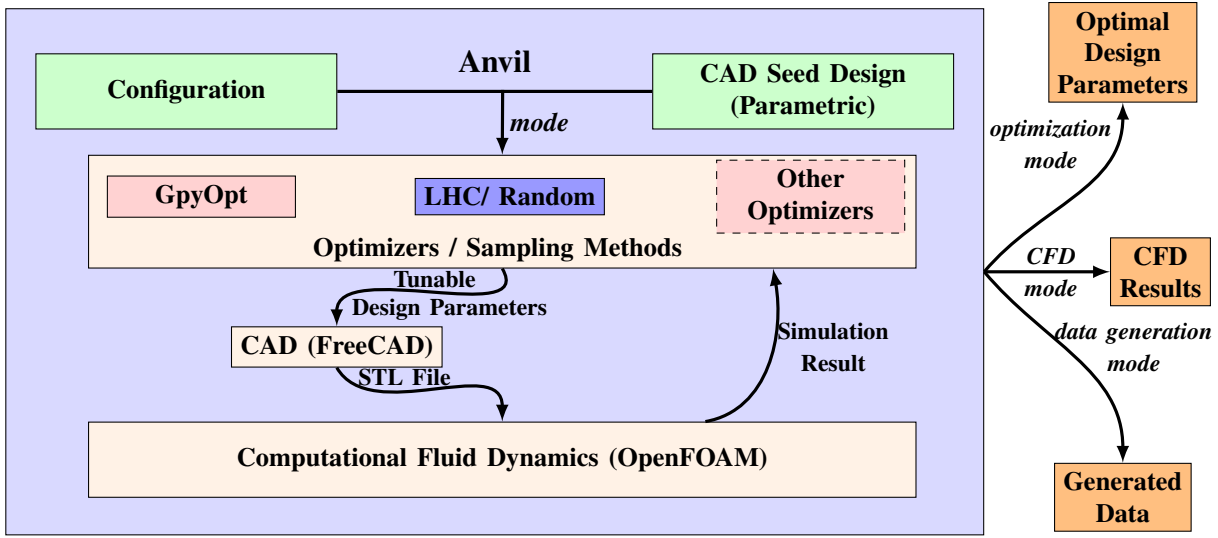
\begin{figure*}[ht!]
    \centering
    \begin{tikzpicture}[scale=0.9]
\node (anvil) [xshift=-6cm, yshift=1cm,fill=blue!15 ,draw=black, thin, minimum height = 7cm, minimum width=13.0cm] at (-1, 0){};

\node (configuration) [align=center, xshift=-11cm, yshift=3.4cm, fill=green!20 ,draw=black, thin, minimum height = 1cm, minimum width = 4cm] at (0, 0){\textbf{Configuration}};

\node (caddesign) [align=center, xshift=-2.8cm, yshift=3.4cm, fill=green!20 ,draw=black, thin, minimum height = 1cm, minimum width = 4cm] at (0, 0){\textbf{CAD Seed Design} \\ \textbf{(Parametric)}};

\node (optimizers) [align=center, xshift=-6.8cm, yshift=1.8cm, fill=orange!10 ,draw=black, thin, minimum height = 1.5cm, minimum width = 11cm] at (0, 0){};

\node (optimizersText) [align=center, xshift=-6.8cm, yshift=1.3cm ,draw=none, thin, minimum height = 2cm, minimum width = 11cm] at (0, 0){
    \textbf{Optimizers / Sampling Methods}
};

\node (gpyopt) [align=center, xshift=-10.8cm, yshift=2cm, fill=pink!70 ,draw=black, thin, minimum width = 2.5cm] at (0, 0){\textbf{GpyOpt}};

\node (lhc) [align=center, xshift=-6.7cm, yshift=2cm, fill=blue!40, draw=black, thin, minimum width = 2.5cm] at (0, 0){\textbf{LHC/ Random}};

\node (other) [align=center, xshift=-2.7cm, yshift=2cm, fill=pink!70 ,draw=black,dashed,thin,minimum width = 2.5cm] at (0, 0){\textbf{Other} \\ \textbf{Optimizers}};

\node (cad) [align=center, xshift=-9cm, yshift=0cm, fill=orange!10 ,draw=black,thin] at (0, 0){\textbf{CAD (FreeCAD)}};

\node (tunableParameters) [align=center, xshift=-7.7cm, yshift=0.9cm, ,draw=none,thin,minimum height = 1.0cm] at (0, 0){\small{\textbf{Tunable}}};

\node (tunableParameters2) [align=center, xshift=-7.5cm, yshift=0.5cm, ,draw=none,thin,minimum height = 1.0cm] at (0, 0){\small{\textbf{Design Parameters}}};

\node (stlFile) [align=center, xshift=-8.1cm, yshift=-0.4cm, ,draw=none,thin,minimum height = 1.0cm] at (0, 0){\small{\textbf{STL File}}};

\node (simulators) [align=center, xshift=-6.8cm, yshift=-1.5cm, fill=orange!10 ,draw=black, thin, minimum height = 1cm, minimum width = 11cm] at (0, 0){\textbf{Computational Fluid Dynamics (OpenFOAM)}};


\node (tunableParameters) [align=center, xshift=-2.4cm, yshift=0cm, ,draw=none,thin,minimum height = 1.0cm] at (0, 0){\small{\textbf{Simulation}}\\\small{\textbf{Result}}};

\node (optdesign) [right of=anvil, align=center, xshift=0cm, yshift=-0.4cm,fill=orange!50 ,draw=black, thin] at (0.5, 4.8){\textbf{Optimal} \\ \textbf{Design} \\ 
\textbf{Parameters}
};

\node (gendata) [right of=anvil, align=center, xshift=0cm, yshift=2.5cm,fill=orange!50 ,draw=black, thin] at (0.5, -5){\textbf{Generated} \\ \textbf{Data}
};

\node (cfd) [right of=anvil, align=center, xshift=0.5cm, yshift=5.5cm,fill=orange!50 ,draw=black, thin] at (0.5, -5){\textbf{CFD} \\ \textbf{Results}
};

\node (anvilText) [below of=anvil, xshift=0cm, yshift=3.8cm, align=center]{\large{\textbf{Anvil}}};
\node (anvilText) [below of=anvil, xshift=0.6cm, yshift=3.0cm, align=center]{\textbf{\textit{mode}}};

\node (datagenmode) [below of=anvil, xshift=8.5cm, yshift=0cm, align=center]{\textbf{\small{\textit{data generation}}} \\ \textbf{\small{\textit{mode}}}};

\node (optmode) [below of=anvil, xshift=7.4cm, yshift=2.7cm, align=center]{\textbf{\small{\textit{optimization}}} \\ \textbf{\small{\textit{mode}}}};

\node (cfdmode) [below of=anvil, xshift=7.4cm, yshift=1.0cm, align=center]{\textbf{\small{\textit{CFD}}} \\ \textbf{\small{\textit{mode}}}};

\draw [->, very thick, -latex] (anvil.east) to [out=50,in=-90] (optdesign.south);
\draw [->, very thick, -latex] (anvil.east) to [out=-50,in=90] (gendata.north);

\draw [-, very thick] (configuration) to (caddesign);

\draw [->, very thick, -latex] (anvil.east) to (cfd);

\draw [->, very thick, -latex] (optimizers.south) to [out=-60,in=90] (cad.north);

\draw [->, very thick, -latex] (-7.55, 3.8) to (optimizers.north);

\draw [->, very thick, -latex] (cad.south) to [out=-60,in=90](simulators.north);

\draw [-, very thick, -latex] ($(simulators.north)+(3,0)$) to [out=10,in=-80] ($(optimizers.south)+(4,0)$);

\end{tikzpicture}
    \caption{Anvil Architecture Diagram. With a configuration JSON file and a parametric CAD seed design, designers can operate Anvil in three modes: (i.) Data Generation, where the CFD results by sampling provided design space can be saved in batches (ii.) CFD, for running CFD on a single parametric design (iii.) Optimization, for searching for optimum design parameters provided a design space and CFD drag, with Bayesian Optimization  }
    \label{fig:anvil_arch}
\end{figure*}

Figure~\ref{fig:anvil_arch} shows the overall software architecture and workflow of \textit{Anvil}.
The significant components shown in the diagram are detailed in the following subsections.

\subsection{User inputs}
In \textit{Anvil}, the user needs to provide two things: first, a configuration JSON file and second, a parametric CAD seed design.  The configuration JSON file contains various parameters, options, or preferences that the user can customize to control how the tool should operate or process the data. By using a parametric CAD seed design, \textit{Anvil} can leverage its parametric capabilities to adjust or adapt the design based on the parameters defined in the CAD model. 

\subsection{Parametric CAD seed design}  
A parametric CAD model provides the flexibility to change the design of a 3D CAD model based on one or more parameters.  The parameters define the shape of the design; this may be viewed as a reduced-order model where the model is controlled by this typically small set of parameters. We use FreeCAD \cite{riegel2016freecad} as our our CAD design tool. To follow the rigorous design practices, we suggest for fully constraining the sketches of CAD design.  To standardize the interaction between the CAD tool (FreeCAD) and Python scripts, we use the \texttt{Spreadsheet} functionality of FreeCAD.
A spreadsheet must be attached to the parametric seed design with the name `Spreadsheet' that consists of all the possible parameters that can be varied during optimization. The experimenter must follow specific rules while designing this parametric CAD design so that \textit{Anvil} can access these parameters and automatically modify the design during experimentation. This rule of writing a spreadsheet is such that all the parameters' variable names with the default values must be provided. These cells in the spreadsheet must be linked with the sketch parameters in the design. By doing this, we do not need to understand the sketch properties and how these sketches interact, and only by interacting with the spreadsheet wrapper, we standardize the interface between Python-based scripts with different FreeCAD designs.

\subsection{STL generation} 
\textit{Anvil} uses the STL file format to exchange the shape generated by a parametric model in FreeCAD to our chosen CFD software. STL (Standard Tessellation Language) is a standard file format used and supported by many other software packages and is widely used for rapid prototyping, 3D printing, and computer-aided manufacturing.
Although the actual standard of the STL file format is dimensionless, FreeCAD assumes that the length unit used in STL files is millimeters (mm). Consequently, all units and parameters in \textit{Anvil} are taken to be consistent with using millimeters as the unit of length. 

\subsection{Computational Fluid Dynamics}
In our problems of interest, it is necessary to evaluate the behavior of a specific solid design immersed in a turbulent fluid.
We assume the fluid to be Newtonian, incompressible, and isothermal.  
Rather than more computationally expensive direct numerical simulation of the governing equations, we adopt a commonly-employed approach of using Reynolds-averaged Navier-Stokes (RANS) paired with a turbulence model.
We employ RANS with the $k$-$\omega$ shear stress transport (SST) model for turbulence physics \cite{menter1992improved}. The $k$-$\omega$ SST model is based on the Boussinesq hypothesis that assumes a relationship between the Reynolds stress tensor and the mean rate-of-strain (deformation) tensor and the turbulent kinetic energy $k$.
To maximize accuracy, the $k$-$\omega$ SST model utilizes the $k$-$\omega$ model \cite{wilcox1998turbulence} near boundaries and the $k$-$\epsilon$ model \cite{launder1983numerical} in the free stream away from the boundary layer.
The $k$-$\omega$ SST model has been successfully validated in various real-world turbulent flow problems \cite{menter1992improved,tide2008comparison,nikiforow2016designing}.
\textit{Anvil} uses OpenFOAM \cite{jasak2007openfoam} as our fluid simulation tool, which provides an implementation of RANS with $k$-$\omega$ SST out of the box.

\subsection{Mesh Generation and auto-meshing}
For volumetric meshing of the fluid portion of the computational domain, we leverage OpenFOAM's built-in meshing tools: \texttt{blockMesh} and \texttt{snappyHexMesh}. 
Meshing is a two steps process: the first step starts with the creation of a castellated 3D parametric volumetric mesh to fill the 3D volume using the \texttt{blockMesh} utility. This volume meshing process uses only hexahedral elements whose number is defined by the parameter in a configuration file. The volumetric mesh is further split and refined in the vicinity of the body surfaces. Once the mesh in the locality of the body is refined and split, cells inside the body shape are removed, keeping the constraint that at least one volumetric region must be bounded inside the domain. This results in separating the body shape from the volume mesh and creating a volume mesh around the surface.  In the next step, refinement of the mesh near the surface layer is performed to generate three-dimensional unstructured or hybrid meshes consisting of hexahedra (hex) and split-hexahedra (split-hex) elements.  Performed with the \texttt{snappyHexMesh} tool, this step removes cells that violate the mesh quality parameters near body surfaces.

For the meshing process, we need to provide the mesh size (level of refinement). The selection of mesh size depends on the size and the shape of the design. Since there is no analytical closed form solution to find the right mesh size, accordingly if an inappropriate mesh size is provided, the simulation fails. To address this, we employ an \textit{auto-meshing} capability, where the mesh size is selected automatically by capturing log errors from the meshing tool and by refining the mesh size accordingly to find the suitable mesh size.

\subsection{Boundary and Initial Conditions}
\label{subsec:BIC}
For all simulations, the appropriate boundary conditions used are the standard Dirichlet boundary conditions. On the left-hand side of the simulation domain, a velocity inlet boundary condition is imposed, while a zero pressure boundary condition is applied on the right-hand side.
The hull surface is subjected to a no-slip boundary condition. The remaining surfaces, including the side wall, symmetry wall, and far-field, are set to have symmetry boundary conditions.
The initial values for the variables $k$ and $\omega$ are determined using the same procedure described in equations \cite{vardhan2023sample}. These initial values are calculated based on the assumption of the Newtonian model of the fluid, which considers the kinematic viscosity to be constant. The user has the option to provide the value of kinematic viscosity based on empirical data in the configuration file. The initial velocity inlet and values of  $k$ and $\omega$ are also entered in the configuration JSON file.

\section{Computational Experiments}
\label{sec:casestudy}
We classify our experiments into three categories based on the different modes in which \textit{Anvil} can be utilized.
\subsection{CFD simulation experiments}
\label{sec:sim_exp}
We first demonstrate how our tool simplifies computational fluid dynamics (CFD) analysis, requiring minimal effort from the user. To conduct a CFD analysis, the designer simply needs to provide an STL file of the design and a configuration file to specify turbulence models, fluid parameters, and other settings.
An example configuration file is available in the \href{https://github.com/symbench/Anvil/blob/main/artifacts/sample_input_config_winged_design.json}{anvil repository}\footnote{\url{https://github.com/symbench/Anvil/blob/main/artifacts/sample_input_config_winged_design.json}}.
We chose to experiment with three different problems: unmanned underwater vehicles (UUVs), land vehicles, and unmanned aerial vehicles (UAVs).
UUVs encompass both autonomous and remotely operated variants. In the design process of UUVs, CFD analysis is crucial for optimizing the hull shape, which significantly influences the vehicle's drag and, consequently, its range and efficiency. For land vehicles, CFD simulation is a valuable tool for analyzing and optimizing the aerodynamic performance. It is employed for various purposes, such as reducing drag, optimizing cooling systems, and studying crosswind stability. In the design of UAVs, CFD is deployed for early design studies and optimization that allow analyzing the aerodynamic performance of UAVs under various flight conditions without the need for physical prototypes or wind tunnel testing. When a CFD simulation reaches steady state, we obtain flow fields on the volumetric mesh that give the user a sense of how fluid flows around a candidate design (shown in Figures \ref{fig:truck:pf}, \ref{fig:truck:vf}, \ref{fig:UAV:pf}, \ref{fig:truck:vf}, \ref{fig:UAV:pf}, and \ref{fig:UAV:vf}). Other physical metrics of interest like drag and lift can be derived using the flow field. Below we present the configuration parameters and CFD flow field results conducted using \textit{Anvil}:

\subsubsection{Unmanned Underwater Vehicle (UUV)}
\begin{figure}[ht]
  \centering
  \begin{minipage}[b]{0.45\textwidth}
  \subfloat[]{
       \includegraphics[width=\textwidth]{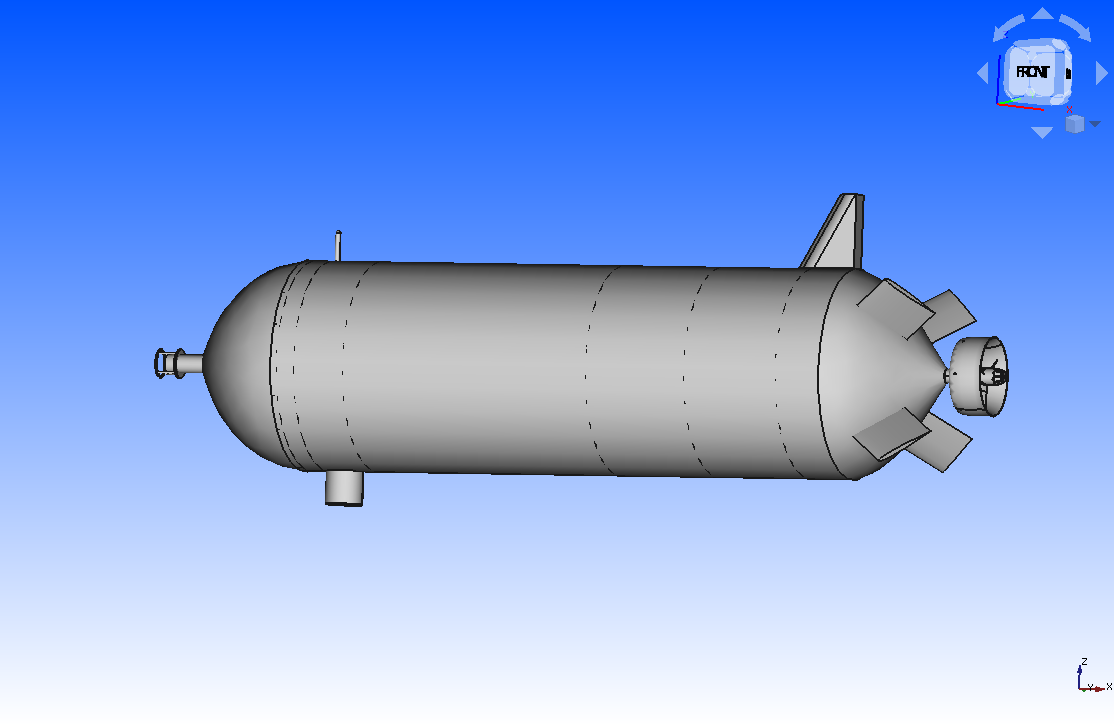}
        \label{fig:UUV:des}
    }
  \end{minipage}
  \begin{minipage}[b]{0.5\textwidth}
    \subfloat[]{
        \includegraphics[width=0.45\textwidth]{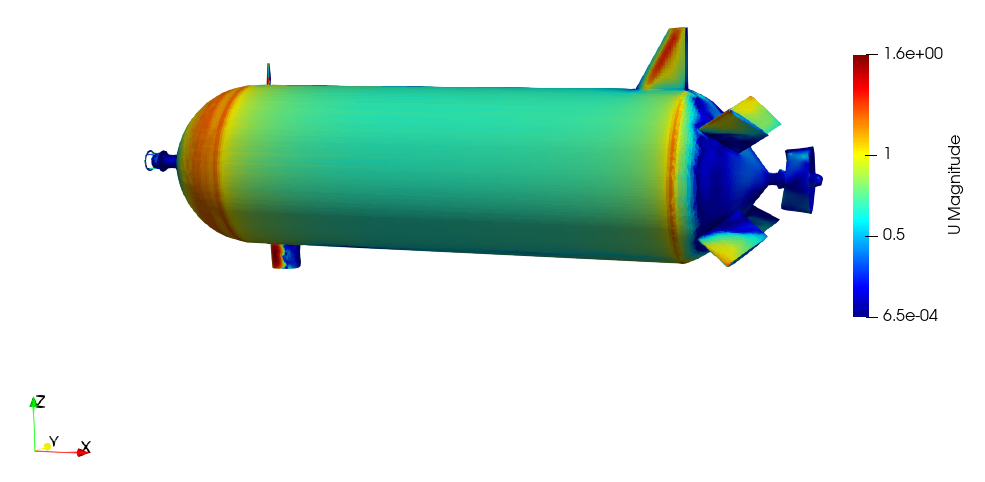}
        \label{fig:UUV:vf}
    }
    \subfloat[]{
        \includegraphics[width=0.45\textwidth]{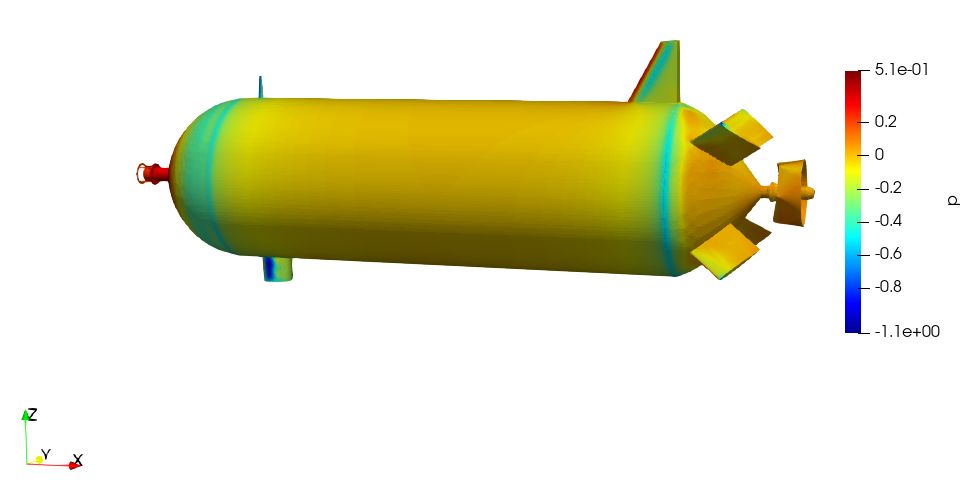}
        \label{fig:UUV:pf}
    }    
\end{minipage}
 \caption{ \protect\subref{fig:UUV:des} The UUV design. Steady state flow at the surface layer of design, with the flow field colored by   \protect\subref{fig:UUV:vf} Velocity field (meters/second), \protect\subref{fig:UUV:pf} Pressure field (Pascals).}
 \label{fig:UUV}
\end{figure}

For this experiment, a UUV design is taken from current research efforts \cite{maroti2022rapid,vardhan2023constrained}. The UUV is simulated at a speed of $2.25$ miles/hr, with fluid properties set to open sea density ($1027.0 Kg/m^3$), dynamic viscosity ($1.789×10^{-5}$ N$-$s/$m^2$), and turbulence intensity of $4\%$, indicating medium turbulence (refer to Figure \ref{fig:UUV}).

\subsubsection{Land vehicle}  
\begin{figure}[ht]
  \centering
  \begin{minipage}[b]{0.45\textwidth}
  \subfloat[]{
       \includegraphics[width=\textwidth]{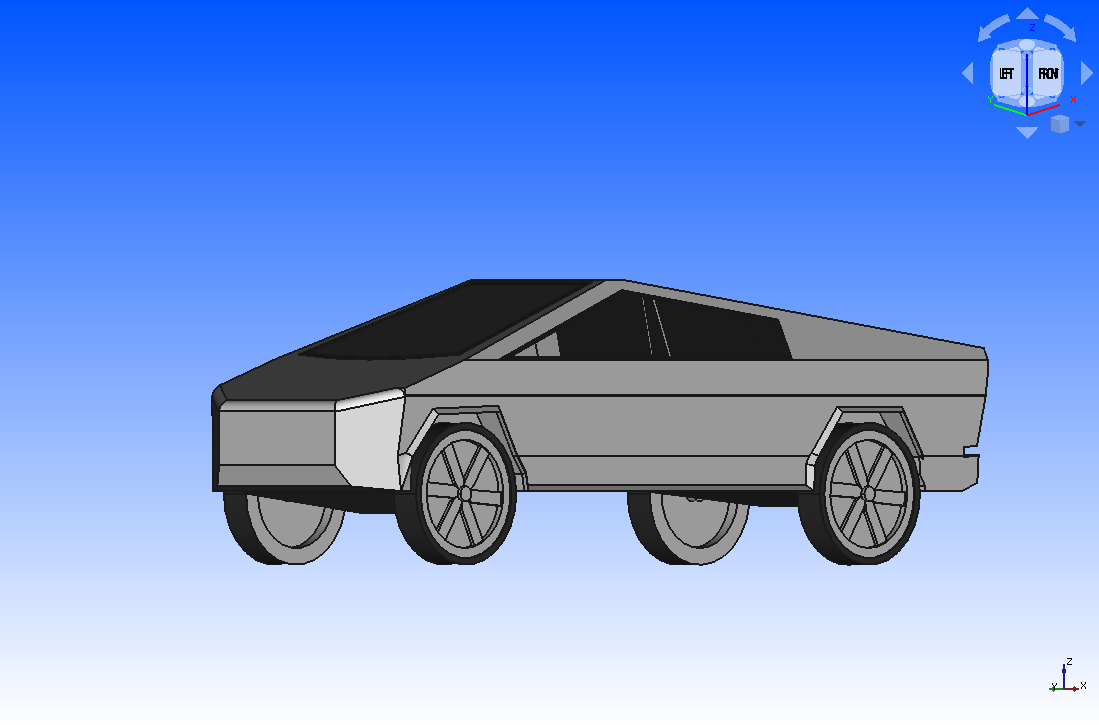}
        \label{fig:truck:des}
    }
  \end{minipage}
  \begin{minipage}[b]{0.5\textwidth}
    \subfloat[]{
        \includegraphics[width=0.45\textwidth]{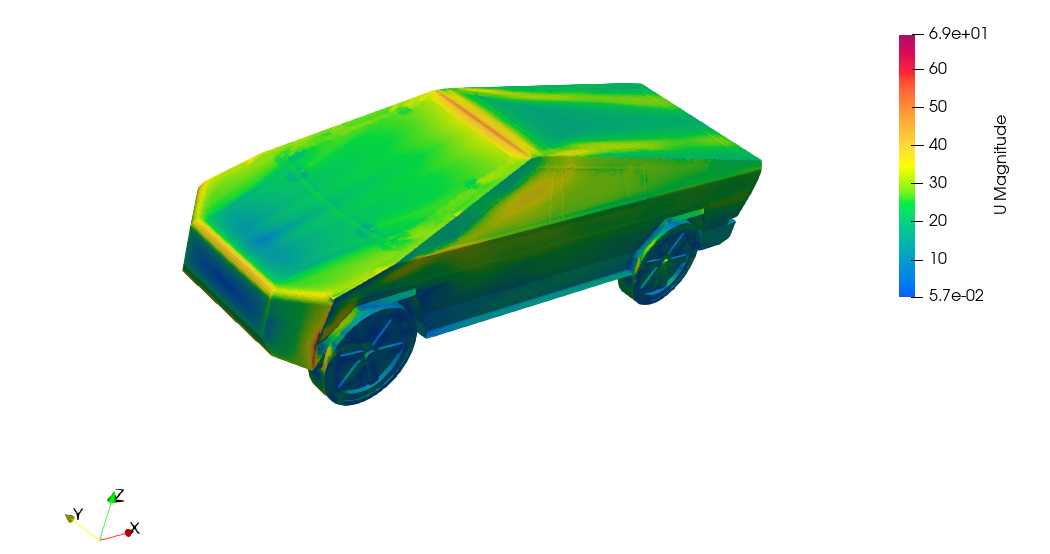}
        \label{fig:truck:vf}
    }
    \subfloat[]{
        \includegraphics[width=0.45\textwidth]{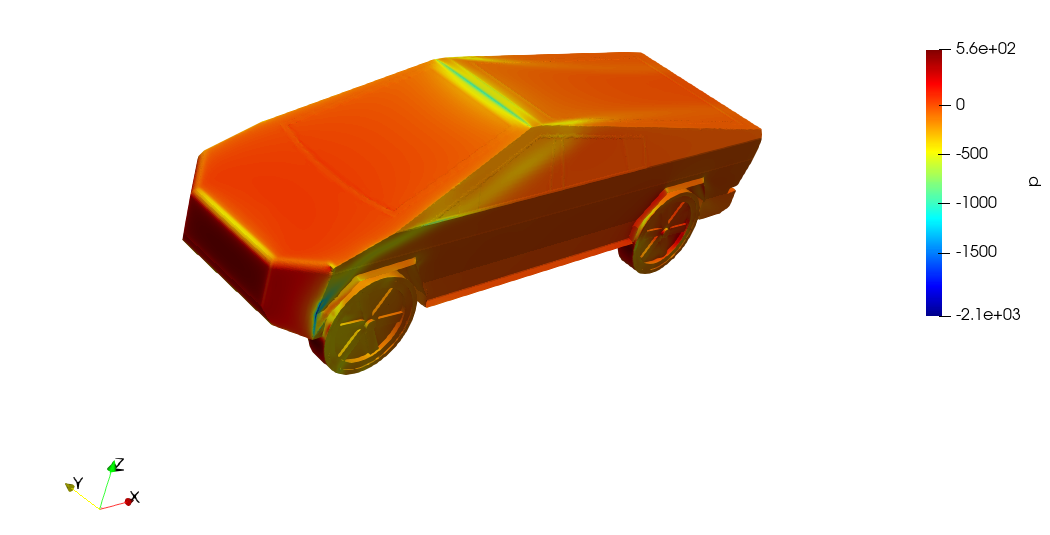}
        \label{fig:truck:pf}
    }
    
  \end{minipage}
 \caption{  \protect\subref{fig:truck:des} The land vehicle design. Steady state flow at the surface layer of design, with the flow field colored by  \protect\subref{fig:truck:vf} Velocity field (meters/second), \protect\subref{fig:truck:pf} Pressure field (Pascals).}
 \label{fig:cybertruck}
\end{figure}

In this experiment, we designed a representative land vehicle and conducted a simulation with the following configuration: the speed of the vehicle was set to be 70 miles/hr, the fluid density is air density on surface i.e. $1.225 Kg/m^3$, dynamic viscosity as $1.789\times10^{-5} N-s/m^2$, and turbulence intensity is at $1\%$ (refer to Figure \ref{fig:cybertruck}).

\begin{figure}[ht]
  \centering
  \begin{minipage}[b]{0.45\textwidth}
  \subfloat[]{
       \includegraphics[width=\textwidth]{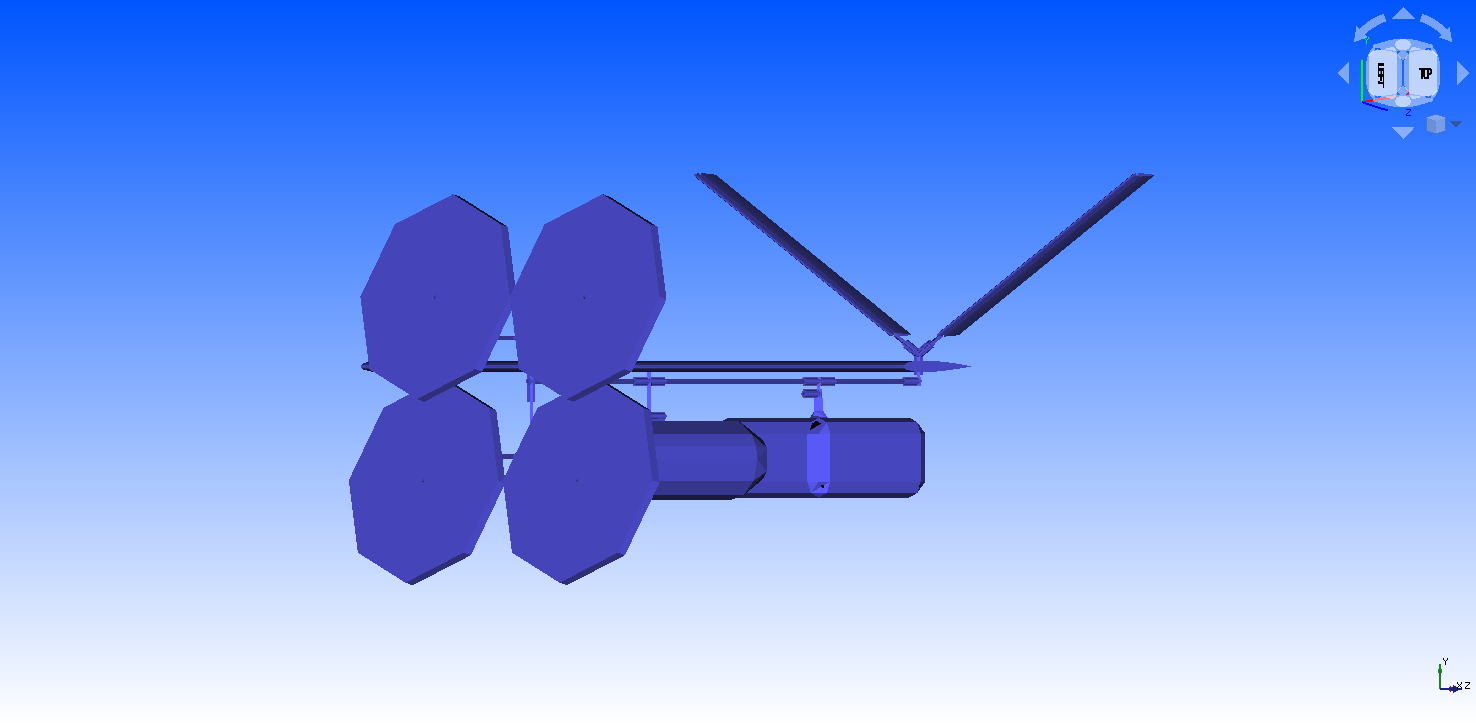}
        \label{fig:UAV:des}
    }
  \end{minipage}
  \begin{minipage}[b]{0.5\textwidth}
    \subfloat[]{
        \includegraphics[width=0.45\textwidth]{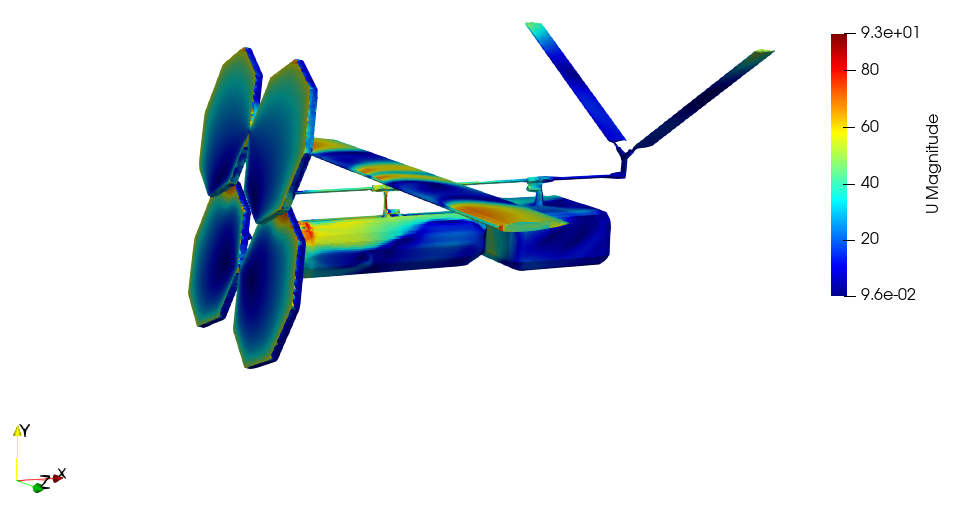}
        \label{fig:UAV:vf}
    }
    \subfloat[]{
        \includegraphics[width=0.45\textwidth]{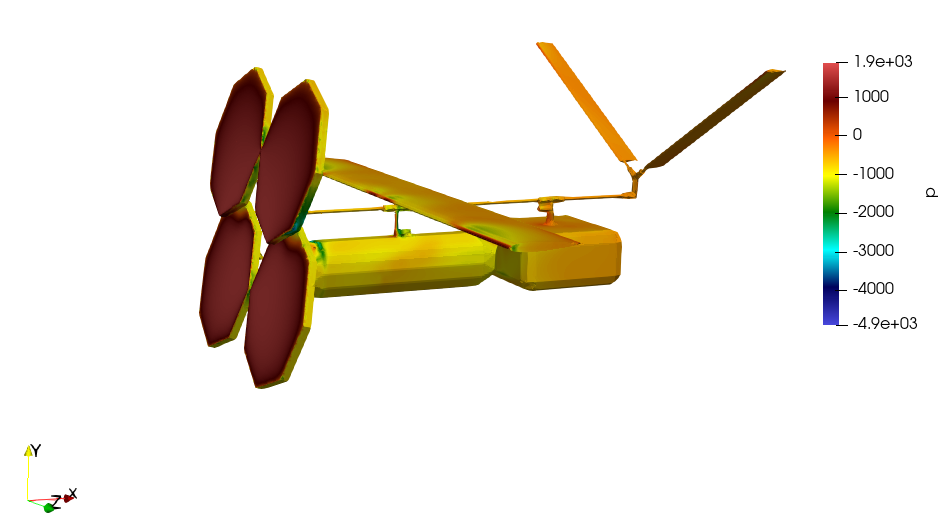}
        \label{fig:UAV:pf}
    }
    
  \end{minipage}
 \caption{  \protect\subref{fig:UAV:des} The UAV design. Steady state flow at the surface layer of design, with the flow field colored by  \protect\subref{fig:UAV:vf} Velocity field (meters/second), \protect\subref{fig:UAV:pf} Pressure field (Pascals).}
 \label{fig:UAV}
\end{figure}

\subsubsection{Unmanned Air Vehicle (UAV)}
Our test UAV design features four propellers positioned at the front, primarily responsible for generating thrust. Additionally, it includes three wings: one dedicated to lift generation, and the other two tasked with controlling roll and yaw. The aircraft is also equipped with a cargo box for storage and a separate compartment housing electronics, batteries, and other essential components. The design is  
simulated at a speed of 50 meters/sec with fluid density is air density $1.225 Kg/m^3$, dynamic viscosity $1.789\times10^{-5} N-s/m^2$, turbulence intensity $1\%$ (refer to Figure \ref{fig:UAV}).

\subsection{Optimization experiment}

\begin{figure}[h]
    \centering
    \includegraphics[width=0.45\textwidth]{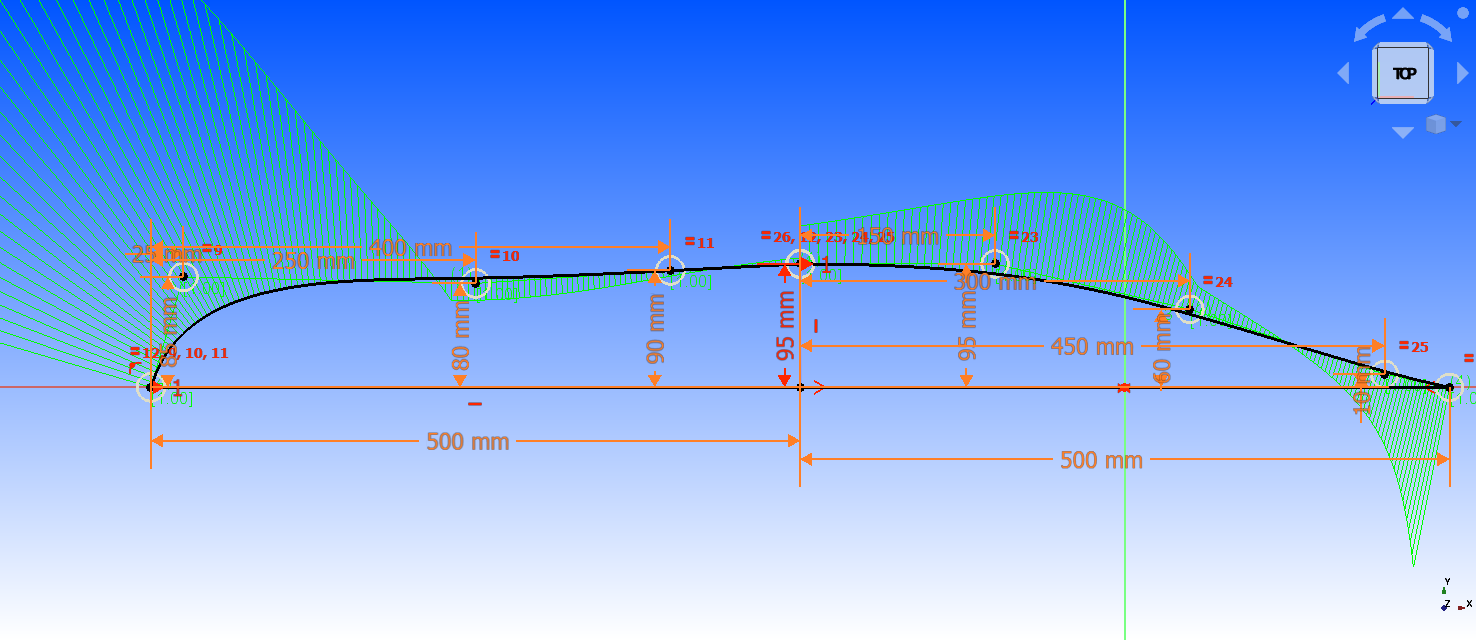}
    \caption{A UUV hull sketch in a CAD environment, with control points are shown in black.}
    \label{fig:uuv_sketch}
\end{figure}

\begin{figure}
    \centering
    \includegraphics[width=0.45\textwidth]{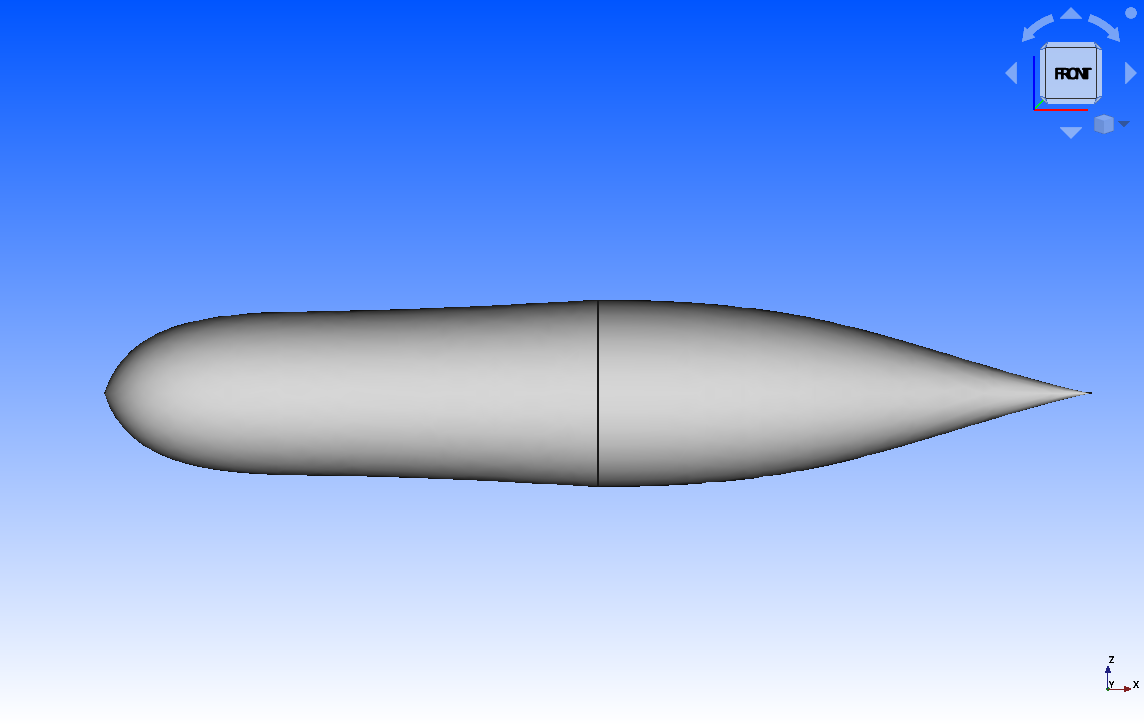}
    \caption{UUV hull shape in a CAD environment after optimization.}
    \label{fig:uuv_hull}
\end{figure}

We also demonstrate the capability of \textit{Anvil} to operate in a design optimization mode.
As a test of this mode, we chose a design optimization problem of an underwater vehicle hull. The problem of finding the optimal hull and its design space is taken from \cite{vardhan2023search}. It considers the $1$ meter length UUV design with $6$ control points to morph the shape of the design (refer to Figure \ref{fig:uuv_sketch}). The design space is in $7$ dimensions: six
control points which are allowed to range between $0$ and $0.2$ meters, and a seventh dimension, the UUV nose length with a search range between $10$ cm - $900$ cm. The tail section is a derived parameter and is defined as $1000$  cm minus the nose length. The optimization goal is to find a hull design with minimal drag. For this we employ sequential Bayesian optimization using with a budget of $50$ iterations. The fluid parameters are set according to $1\%$ turbulence intensity and inlet velocity of $2.5$ meter/sec. 
The details of the Bayesian optimization approach can be found in \cite{vardhan2023sample,frazier2018tutorial}. The hyper-parameters for Bayesian optimization and optimization goal and budgets are set in an \textit{Anvil} configuration file (we chose Lower Confidence Bound (LCB) as our acquisition function). The mixture of radial basis function (RBF) kernel with small white noise is used as the learning kernel for multivariate Gaussian Process regression. The optimal design found after exhaustion of budget is shown in Figure \ref{fig:uuv_hull}.

\subsection{Data generation experiment}
\textit{Anvil} can also be used to generate designs sampled from a design space.
For our test problem, we made a parametric design that represents a winged air vehicle.  In the design of this aerospace vehicle body, there are three main components: a hemispherical nose, a conical tail, and a cylindrical body. The wings of the body are another crucial component, and for symbolic representation, we deploy rectangular wings. Various parameters are considered that can alter the vehicle's shape, including the length of the nose section (denoted as \(nose\_radius\)), the length of the body section (or \(fuselage\_length\)), the tail section's length (referred to as \(tail\_length\)), the thickness of the wing (\(thickness\_wing\)), the span of one wing (\(half\_span\)), and the chord length of the wing. The range of the design space for these parameters is shown in Table \ref{tab:ds_winged}.

\begin{table}
\centering
\normalsize
\captionsetup{justification=centering,format=plain,font=small, labelfont=bf}
\begin{adjustbox}{width=\columnwidth,center}
\begin{tabular}{lccc} 
\toprule
\textbf{Parameter} & \hfil \textbf{Symbol} & \hfil \textbf{ Minimum } & \hfil \textbf{ Maximum } \\
\midrule
Length of nose section & $nose\_radius$ & $100$ mm & $800$ mm  \\
Length of body section  & $fuselage\_length$ & $100$ mm & $800$ mm \\
Length of tail section & $tail\_length$ & $100$ mm & $800$ mm \\
Thickness of wing  & $thickness\_wing$ & $5$ mm & $50$ mm  \\
Span of one wing & $half\_span$ & $50$ mm & $200$ mm\\
Chord length of wing  & $chord$ & $50$ mm & $200$ mm \\
\bottomrule
\end{tabular}
\end{adjustbox}
\vspace{0mm}
\caption{Design Space: Range of design space parameters for data generation.}
\label{tab:ds_winged}
\end{table}

\begin{figure}[h]
    \centering
    \includegraphics[width=0.45\textwidth]{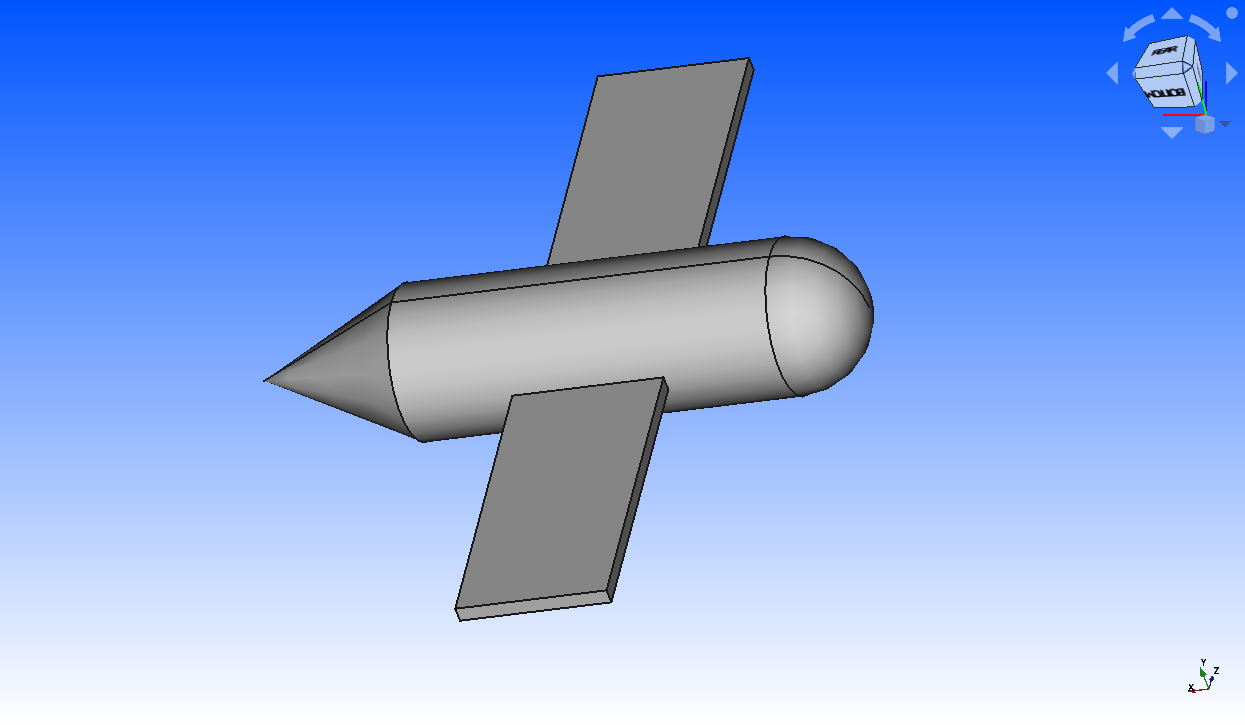}
    \caption{The winged air vehicle design in CAD environment}
    \label{fig:wing_des}
\end{figure}
The budget (the number of labeled data points to generate) and sampling methods can both be specified in an \textit{Anvil} configuration file. Currently \textit{Anvil} supports uniform random sampling, Latin hypercube based on minimizing the maximum correlation, and Latin hypercube basewd on maximizing the minimum
distance between points. For this experiment, we chose uniform random sampling from the design space and a budget of $100$ samples.

\section{Limitations}
\label{sec:limitation}
\textit{Anvil} is capable of addressing the shape optimization problem for a wide range of air and water-based designs, provided they operate at velocities below the transonic range. 
At low Mach numbers, air can be considered nearly incompressible, or isentropic, because the changes in flow variables are small and gradual. The assumption of incompressibility is generally safe up to $1$ Mach, or until the flow becomes transonic \cite{NASAweb}. A second limitation of Anvil is that, despite its adaptive mesh size selection, it does not perform a convergence test of the outcomes using different refined mesh sizes. Finally, the tool is limited to handling design optimization problems with up to 20 dimensions due to the constraints of the underlying optimization algorithm, Bayesian Optimization \cite{frazier2018tutorial}.
\section{Conclusions \& Future Work}
\label{sec:conclusion}
Our work presents \textit{Anvil}, an open-source integrated CAD-CFD tool through the combination of FreeCAD, OpenFOAM, Bayesian optimization, and other algorithms. \textit{Anvil} proves to be a versatile scientific machine learning tool, capable of handling data generation, CFD evaluation, and shape optimization for a wide range of subsonic flow conditions, as demonstrated in various use cases. In future work, efforts will be directed towards extending \textit{Anvil}'s capabilities to handle transonic and supersonic flow conditions. Additionally, we plan to enhance the tool's functionality by incorporating convergence testing with varied mesh sizes and expanding its capacity to tackle optimization problems in higher dimensions.
\section{Acknowledgments}
This work is funded by  Defense Advanced Research Projects Agency (DARPA)’s Symbiotic Design for CPS project and by the Air Force Research Laboratory (FA8750-20-C-0537).
\bibliographystyle{IEEEtran}
\bibliography{t08Bibliography}

\end{document}